\documentclass[aps,prl,twocolumn,showpacs,preprintnumbers,amsmath,amssymb,superscriptaddress]{revtex4}%

\usepackage{graphicx}
\usepackage{dcolumn}
\usepackage{bm}

\begin{document}
\preprint{PREPRINT (\today)}

\title{Muon-Spin-Rotation Measurements of the Penetration Depth in the
Infinite-Layer Electron-Doped Cuprate Superconductor Sr$_{0.9}$La$_{0.1}$CuO$_{2}$}
\author{A.~Shengelaya} 
\affiliation{Physik-Institut der Universit\"{a}t
Z\"{u}rich, Winterthurerstrasse 190, CH-8057 Z\"{u}rich, Switzerland}
\author{R.~Khasanov} 
\affiliation{Physik-Institut der Universit\"{a}t
Z\"{u}rich, Winterthurerstrasse 190, CH-8057 Z\"{u}rich, Switzerland}
\affiliation{ Laboratory for Neutron Scattering, ETH Z\"urich and
Paul Scherrer Institut, CH-5232 Villigen PSI, Switzerland}
\affiliation{DPMC, Universit\'e de Gen\`eve, 24 Quai
Ernest-Ansermet, 1211 Gen\`eve 4, Switzerland}

\author{D. G.~Eshchenko} 
\affiliation{Physik-Institut der Universit\"{a}t
Z\"{u}rich, Winterthurerstrasse 190, CH-8057 Z\"{u}rich, Switzerland}
\affiliation{Paul Scherrer Institut, CH-5232 Villigen PSI, Switzerland}
\author{D. Di~Castro} 
\affiliation{Physik-Institut der Universit\"{a}t
Z\"{u}rich, Winterthurerstrasse 190, CH-8057 Z\"{u}rich, Switzerland}
\author{I. M.~Savi\'c} 
\affiliation{Faculty of Physics, University of Belgrade, 11001 Belgrade,
Serbia and Montenegro}
\author{M.S.~Park} 
\affiliation{National Creative Research Initiative Center for
Superconductivity and Department of Physics, Pohang University of
Science and Technology, Pohang 790-784, Republic of Korea}
\author{K.H.~Kim} 
\affiliation{National Creative Research Initiative Center for
Superconductivity and Department of Physics, Pohang University of
Science and Technology, Pohang 790-784, Republic of Korea}
\author{Sung-Ik~Lee} 
\affiliation{National Creative Research Initiative Center for
Superconductivity and Department of Physics, Pohang University of
Science and Technology, Pohang 790-784, Republic of Korea}
\author{K.A.~M\"uller} 
\affiliation{Physik-Institut der Universit\"{a}t Z\"{u}rich,
Winterthurerstrasse 190, CH-8057 Z\"{u}rich, Switzerland}
\author{H.~Keller} 
\affiliation{Physik-Institut der Universit\"{a}t
Z\"{u}rich, Winterthurerstrasse 190, CH-8057 Z\"{u}rich, Switzerland}
%

\begin{abstract}
Muon spin rotation ($\mu$SR) measurements of the in-plane penetration 
depth $\lambda_{ab}$ have been performed in the electron-doped infinite 
layer high-$T_{c}$ superconductor (HTS) Sr$_{0.9}$La$_{0.1}$CuO$_{2}$.
Absence of the magnetic rare-earth ions in this compound allowed to
measure for the first time the absolute value of $\lambda_{ab}(0)$ in
electron-doped HTS using $\mu$SR. We found $\lambda_{ab}(0)$=116(2) nm.  
The zero-temperature depolarization rate
$\sigma(0)\propto1/\lambda^{2}_{ab}(0)$=4.6(1) $\mu$s$^{-1}$ is more
than four times higher than expected from the Uemura line.  Therefore
this electron-doped HTS does not follow the Uemura relation found for
hole-doped HTS.
    
\end{abstract}

\maketitle
The high-$T_{c}$ cuprate superconductors are obtained by doping holes
or electrons into the antiferromagnetic (AF) insulating state.  Both electron
and hole-doped cuprates share a common building block, namely the
copper-oxygen plane and one would expect that the same pairing mechanism 
is applicable. There are a number
of important differences, however, between the generic phase diagrams
of the electron-doped and hole-doped materials.  In order to elucidate
the mechanism of high-$T_{c}$ superconductivity (HTS), it is very
important to clarify the origin of similarities and
differences between hole-doped ($p$-type) and electron-doped ($n$-type)
cuprates.

The magnetic field penetration depth $\lambda$ is one of the fundamental 
lengths of a superconductor, related to the superfluid phase 
stiffness $\rho_{s}\propto 1/\lambda^{2}$, or what is often referred 
to as superfluid density $n_{s}/m^{*}\propto 1/\lambda^{2}$
(superconducting carrier 
concentration n$_{s}$ divided by the effective mass m$^{*}$). 
Accurate and precise measurements of the absolute value of $\lambda(T 
\rightarrow 0)$ are very important for understanding superconductivity in cuprates.
The muon-spin-rotation ($\mu$SR) technique provides a powerful tool to
measure $\lambda$ in type II superconductors.
Detailed $\mu$SR investigations of polycrystalline
HTS have demonstrated that $\lambda$ can be obtained from the muon
spin depolarization rate $\sigma(T) \propto 1/\lambda^{2}(T)$, which probes
the second moment of the magnetic field distribution in the mixed
state \cite{Review}.  One of the most interesting result of $\mu$SR
investigations in HTS is a remarkable proportionality between $T_{c}$
and the zero-temperature depolarization rate $\sigma(0)\propto
1/\lambda^{2}(0)$ for a wide range of $p$-type underdoped HTS 
(so-called Uemura line) \cite{Uemura1,Uemura2}.  
This observation indicates that the
superfluid density is an important quantity which determines $T_{c}$
in HTS. This is not expected in conventional BCS theory and therefore
the Uemura relation has an important implication for the physics of
HTS \cite{Millis}.

Unfortunately, it is not known up to now whether the $n$-type cuprates
also obey the Uemura relation.  The large dynamic relaxation due to
rare-earth magnetic moments in $n$-type cuprates
R$_{2-x}$Ce$_{x}$CuO$_{4-\delta}$ (R= Nd, Sm, Pr) with so called
$T'$-structure prevented the determination of $\sigma(0)$ in $\mu$SR
experiments \cite{Luke}.  Because of this problem, other
techniques like microwave surface impedance and magnetization were used
to determine the penetration depth in $n$-type cuprates.  However, it
is difficult to determine the absolute value of $\lambda$ with these
experiments and the reported values vary in a very wide range from 100
to 300 nm even for optimally doped samples.  Therefore there is no
consensus about the penetration depth value for $n$-type cuprates. 
Another difficulty concerns the quality of the samples.  A
long-standing mystery for the $T'$-structure $n$-type cuprates is the
effect of an oxygen reducing procedure.  Superconductivity shows up
only when a minute amount ($\Delta y\approx$ 0.02) of interstitial
oxygens are removed by the reducing procedure \cite{Takagi}.  The role
of the tiny amount of interstitial oxygen is not clear up to now.  The
control of the oxygen content requires rather extreme conditions, such
as temperatures as high as 850$^{\circ}$-950$^{\circ}$C in Ar, which
is not far below the sintering temperature.  Therefore the control of
the sample quality and reproducibility becomes a serious problem.

There exists another class of $n$-type cuprates (Sr,Ln)CuO$_{2}$
(Ln=La, Sm, Nd, Gd) with so-called infinite-layer structure
\cite{Siegrist,Smith}.  The $n$-type infinite-layer superconductors
(ILS) have several merits.  First, the simplest crystal structure
among all HTS consisting of an infinite stacking of CuO$_{2}$ planes
and (Sr, Ln) layers.  The charge reservoir block commonly present
in cuprates does not exist in the infinite-layer structure. 
Second, the stoichiometric oxygen content without vacancies or
interstitial oxygen \cite{Jorgensen}.  Third, $n$-type ILS have much
higher $T_{c}\simeq$ 43 K compared to the $n$-type cuprates with
$T'$-structure $T_{c}\simeq$ 25 K. Although $n$-type ILS have
existed for quite a while, not many studies of their physical
properties were performed because of the lack of high-quality samples
with a complete superconducting volume.  Recently, high-quality
$n$-type ILS samples of Sr$_{0.9}$La$_{0.1}$CuO$_{2}$ with a sharp
superconducting transition $T_{c}\simeq$ 43 K were synthesized by 
using a cubic multi-anvil press \cite{Jung}.

In this letter we report studies of the penetration depth $\lambda$ in
Sr$_{0.9}$La$_{0.1}$CuO$_{2}$ ILS using the transverse-field (TF) $\mu$SR
technique.  We confirmed microscopically that this compound is a bulk
superconductor.  Because of the absence of magnetic
rare-earth ions, it was possible to measure the penetration depth in $n$-type
HTS for the first time using $\mu$SR, yielding $\lambda_{ab}(0)$=116(2) nm.
The zero-temperature depolarization rate $\sigma(0)$=4.6(1) $\mu$s$^{-1}$ 
is more than four times larger than expected from the Uemura plot.  
This shows that the $n$-type ILS does not follow the Uemura relation 
established in $p$-type HTS.

The polycrystalline samples Sr$_{0.9}$La$_{0.1}$CuO$_{2}$ (SLCO) for
this study were prepared with the high-pressure technique using a
cubic multi-anvil press \cite{Jung}.  Magnetization measurements showed
a single sharp superconducting transition at
$T_{c}\simeq$ 43 K and the saturation of the susceptibility at low
temperatures indicating the good sample quality.
The $\mu$SR measurements were performed 
at the Paul Scherrer Institute (PSI, Switzerland) using low-momentum muons
(29 MeV/c). A detailed discussion of
the TF-$\mu$SR technique is given in \cite{Pumpin} where details of
the application of the technique to the determination of $\lambda$ can
be found.  

Fig. 1(a) shows TF-$\mu$SR muon-spin precession signals in an applied 
field of 600 mT above and below $T_{c}$. For visualization purposes 
the apparent precession frequencies are modified from the actual 
precession frequencies by the use of a rotating reference frame. 
In the normal state above 
$T_{c}$, the oscillation shows a small relaxation due to random local 
fields from nuclear magnetic moments. Below $T_{c}$, the relaxation rate 
strongly increases due to the inhomogeneous field distribution of the 
flux line lattice. It is well known that in $n$-type cuprates there is a 
competition between the antiferromagnetically ordered state and 
superconductivity  \cite{Luke}. The static magnetism, if present, could 
enhance the muon depolarization rate and  falsify the interpretation 
of the TF-$\mu$SR results.  We have therefore carried out zero-field 
(ZF) $\mu$SR experiments to determine whether such static 
magnetism exists in SLCO.  Typical ZF-$\mu$SR spectra are shown in Fig. 1(b) 
for temperatures above and below $T_{c}$.  The ZF relaxation is 
exponential with a small relaxation rate 0.149(4) $\mu$s$^{-1}$ and 
0.184(5) $\mu$s$^{-1}$ at 50 and 2.5 K, respectively.  Thus, there is 
no evidence for the static magnetism in SLCO down to 2.5 K. Moreover, 
the ZF relaxation rate is small and changes very little between 50 
and 2.5 K. Therefore, the increase in TF relaxation rate below  $T_{c}$ 
is attributed entirely to the vortex lattice.

\begin{figure}[htb]
\includegraphics[width=0.9\linewidth]{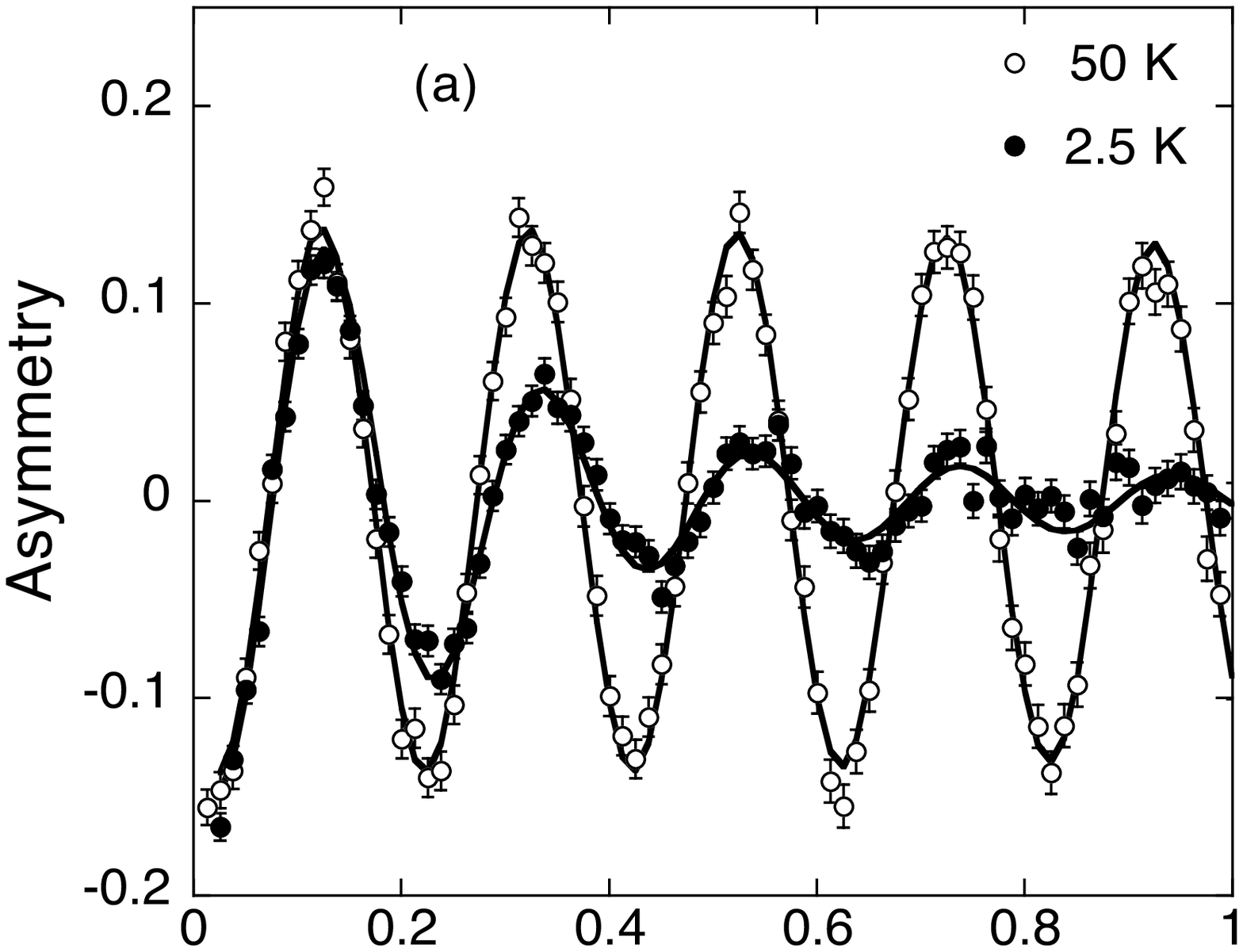}
\includegraphics[width=0.9\linewidth]{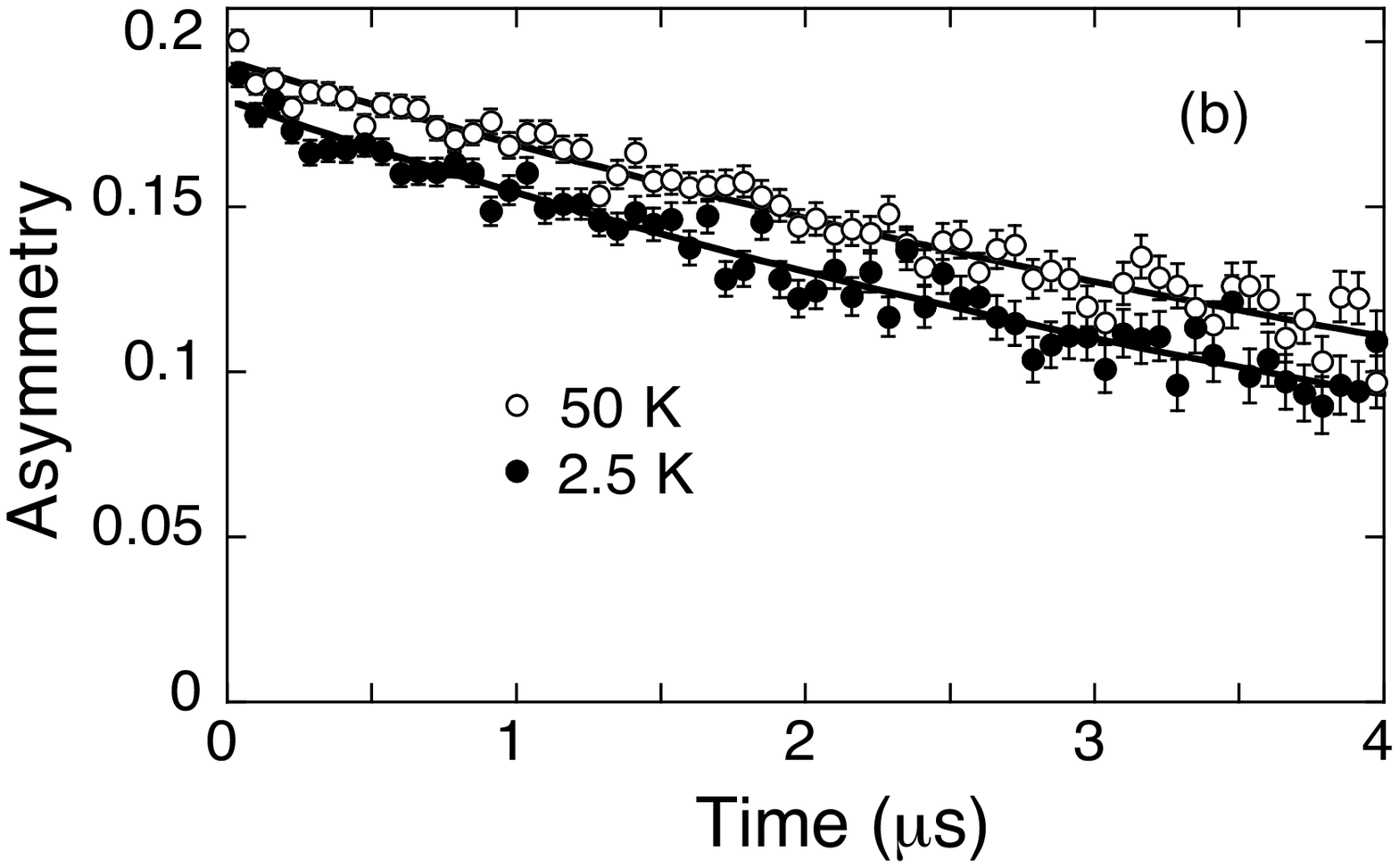}
\caption{ (a) TF-$\mu$SR spectra in  Sr$_{0.9}$La$_{0.1}$CuO$_{2}$ 
in a magnetic field of 0.6 T at temperatures T=50 K (above T$_{c}$) 
and T=2.5 K (below T$_{c}$). Solid lines represent fits using Eq. (1).
(b) ZF-$\mu$SR spectra in Sr$_{0.9}$La$_{0.1}$CuO$_{2}$  at 50 K and 2.5 K. 
Solid lines show fits using an exponential function.}
\label{Fig.1}
\end{figure}

Detailed $\mu$SR experiments in polycrystalline HTS
have shown that the internal field distribution in the mixed state 
can be well approximated by a Gaussian distribution \cite{Pumpin}.
We used two Gaussian model for analyzing our asymmetry time spectra:
\begin{equation} 
A(t)=\sum_{i=1}^{2} A_{i}exp(-\sigma^{2}_{i}t^{2}/2)cos(2\pi\gamma 
B_{i}t+\varphi)
\end{equation}
where $A_{i}$ represent the asymmetries of the two components, 
$\sigma_{i}$ the muon depolarization rates, $B_{i}$  the 
local magnetic fields at the muon sites, $\gamma$=135.5 MHz/T is the 
muon gyromagnetic ratio and $\varphi$ the initial 
phase. The solid lines in Fig.1 (a) show the best fits to Eq. (1). The 
fit is statistically satisfactory ($\chi^2$ criterion), as can be seen qualitatively in 
Fig.1(a). 

Analysis of the asymmetry time spectra showed that below $T_{c}$ in the present SLCO 
sample more than 80 \% of the muons stop in the superconducting 
regions (first component). In this regions the internal magnetic 
field is smaller than the external one because of the diamagnetic screening and the 
depolarization rate is much higher than in the normal state because of the flux-line lattice formation.  
The rest 20 \% of the muons (second component) oscillate with a frequency
nearly equal to that corresponding to the applied magnetic field with
a much smaller depolarization rate.  This signal is most probably
coming from the muons stopping in the nonsuperconducting grain
boundaries and other defects in the structure and is often observed in
polycrystalline HTS \cite{Lichti}.  As already mentioned samples of ILS prepared so 
far suffer from the small volume fraction of the superconducting phase.  
As a real space microscopic
probe $\mu$SR can distinguish between the superconducting and
nonsuperconducting phases and determine their relative volume
fractions.  The present $\mu$SR measurements provide
{\em microscopic} evidence for the excellent quality of the SLCO
ILS prepared with the cubic multi-anvil press technique \cite{Jung}.

In polycrystalline samples the effective penetration depth
$\lambda_{eff}$ (powder average) can be extracted from the $\mu$SR
depolarization rate $\sigma \sim \lambda_{eff}^{-2}$.  It was shown
\cite{Gunn,Smilga} that in polycrystalline samples of highly
anisotropic systems such as the HTS ($\gamma$=$\lambda_{c}/
\lambda_{ab}>5$), $\lambda_{eff}$ is dominated by the shorter
penetration depth $\lambda_{ab}$ and $\lambda_{eff}=1.31\lambda_{ab}$.
Recent magnetization measurements in grain-aligned
SLCO showed a rather high anisotropy value $\gamma$=9 \cite{Kim1,Kim2}. 
Therefore the measured $\lambda_{eff}$ is solely determined by the
in-plane penetration depth $\lambda_{ab}$.

The relation between $\sigma$ and $\lambda_{ab}$ is only valid for
high magnetic fields ($B_{ext}>2\mu_{0}H_{c1}$), when the separation 
between vortices is smaller than $\lambda$. In this case, according to 
the London model $\sigma$ is field independent \cite{Brandt}.  
To check for this, we measured $\sigma$ as
a function of the applied field at T=10 K. Each point was
obtained by field-cooling the sample from above T$_{c}$ to 10 K. The inset of Fig.2 
shows that $\sigma$ strongly increases with 
increasing magnetic field up to $B_{ext}\simeq$50 mT and above 50 mT 
changes very little with magnetic field. Such a behavior is expected
within the London model and is typical for polycrystalline HTS \cite{Pumpin}. 
It can be seen that above 50 mT $\sigma(B)$ shows a tendency of 
gradual decrease with increasing field. This can be due to the 
increase of $\lambda$ with magnetic field due to the anisotropic order 
parameter and the associated nonlinear effect due to the Doppler shift 
of the quasiparticles in the nodal region \cite{Volovik,Sonier}. However, the 
field range in our experiment is too narrow to discuss this in more 
detail. Based on the $\sigma(B)$ measurements,
we studied the temperature dependence of $\sigma$ in a magnetic field
of 600 mT (the largest available field of the GPS spectrometer at
PSI).  We choose the highest magnetic field because at higher fields
the enhanced vortex-vortex interaction helps to
maintain the long-range order of the vortex lattice, which is
important for the determination of $\lambda$.

\begin{figure}[htb]
\includegraphics[width=1.0\linewidth]{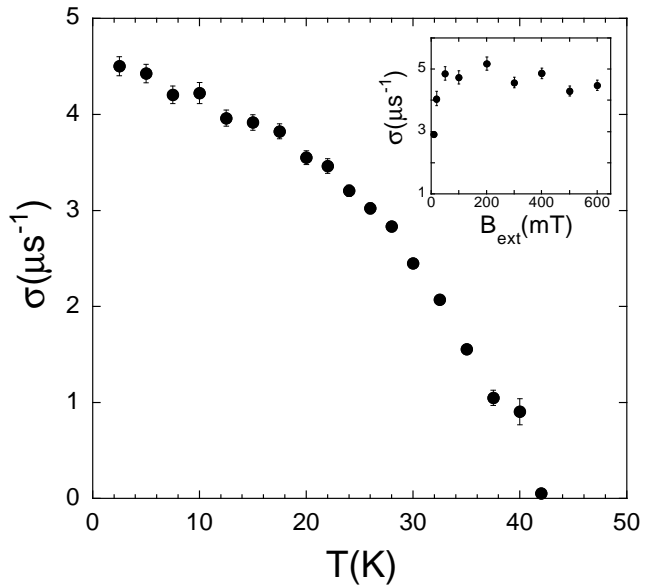}
\caption{Temperature dependence of the $\mu$SR depolarization rate 
$\sigma(T)$ of  Sr$_{0.9}$La$_{0.1}$CuO$_{2}$. Inset: depolarization 
rate as a function of the external magnetic field B$_{ext}$ at 10 K.}
\label{Fig.2}
\end{figure}

Fig.  2 shows the temperature dependence of the muon 
spin depolarization rate $\sigma(T)$ at $B_{ext}$=0.6 T. The values of 
$\sigma(T)$ were derived after subtraction of the small normal state 
temperature-independent depolarization rate originating from the 
copper nuclear moments 
($\sigma(T)^{2}$=$\sigma_{1}^{2}$-$\sigma_{norm}^{2}$).
From the data in Fig. 2 extrapolated to 0 K we obtain the value 
$\sigma(0)$=4.6(1) $\mu$s$^{-1}$ 
which corresponds to  $\lambda_{ab}(0)$=116(2) nm. The
value $\sigma(0)$= 4.6(1) $\mu$s$^{-1}$ is one of the highest among all
HTS. Fig.  3 shows T$_{c}$ plotted versus $\sigma(0)$ (Uemura plot
\cite{Uemura1,Uemura2}) for $p$-type cuprates, including the present
result for $n$-type SLCO. One can see that SLCO strongly deviates from
the Uemura line.  It is interesting to consider the situation in
$n$-type HTS with the $T'$-structure.  As we already mentioned it was
not possible to extract the value of $\lambda$ in this type of
compounds with $\mu$SR, and most of the experiments were performed by
means of microwave surface impedance technique which yielded very
controversial results due to the difficulty to extract the absolute
values of $\lambda$.  There are however two studies of $\lambda(0)$ in
Nd$_{1.85}$Ce$_{0.15}$CuO$_{4}$ (NCCO) single crystals by means of
magnetization \cite{Nugroho} and infrared optics \cite{Homes}.  We
included in Fig.  3 these $\lambda(0)$ values converted to
$\sigma(0)$.  It is seen that similar to ILS
SLCO, $n$-type NCCO with the $T'$-structure strongly deviates from the Uemura
line.  This was also pointed out by Homes {\em et al.} from the
optical measurements \cite{Homes}.  Based on the presented results one
can conclude that $n$-type HTS do not follow the Uemura relation
established in $p$-type HTS.

\begin{figure}[htb]
\includegraphics[width=1.0\linewidth]{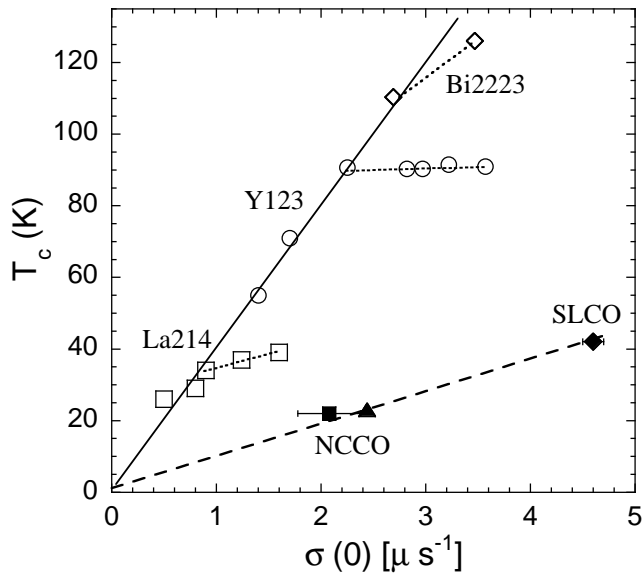}
\caption{T$_{c}$ versus $\sigma(0)$ for $p$- and $n$-type cuprate superconductors.
Open symbols represent data taken from Uemura {\em et al.} 
for various $p$-type HTS \cite{Uemura2}. The solid line is the universal Uemura line
for different underdoped $p$-type HTS.  Dotted lines
represent deviation from the Uemura relation near optimal doping. 
Solid diamond shows the Sr$_{0.9}$La$_{0.1}$CuO$_{2}$ (SLCO) data obtained in
the present work.  Solid square and triangle show the data for
Nd$_{1.85}$Ce$_{0.15}$CuO$_{4-\delta}$ (NCCO) single crystals obtained
from magnetization \cite{Nugroho} and optical \cite{Homes} 
measurements, respectively.  Dashed line represents a tentative
T$_{c}$ vs.  $\sigma(0)$ relation for $n$-type cuprates.}
\label{Fig.3}
\end{figure}

There are several important differences
between the normal-state properties of the $p$- and $n$-type cuprates. 
The $p$-type materials show $T$-linear in-plane electrical resistivity
\cite{Takagi1} and incommensurate magnetic fluctuations \cite{Yamada1}
whereas the $T'$ $n$-type materials show a $T^{2}$ dependence of the
in-plane resistivity \cite{Hagen} and commensurate magnetic
fluctuations \cite{Yamada2}.  Recent NMR experiments in $n$-type
cuprates found no evidence of the pseudogap in contrast to $p$-type
materials \cite{Williams,Zheng}. Present results show that the
differences between the $p$- and $n$-type cuprates extend also to the
superconducting state.  Namely, we observed that in $n$-type cuprates
the superfluid density $n_{s}/m^{*}$ is more than four times larger 
compared to $p$-type cuprates with the same $T_{c}$.

Finally let us comment the temperature dependence $\sigma(T)$ presented in
Fig.  2.  One can see that at low temperatures (below $\sim$15 K)
$\sigma(T)$ is not constant and instead follows the linear
temperature dependence.  Usually a linear low-temperature behavior
of $\sigma(T)$ is taken as an indication for a $d$-wave gap function
with line nodes \cite{Scalapino}.  However, experience with $p$-type cuprates
showed that the single crystals are required for conclusive determination
of the intrinsic temperature dependence of $\sigma$ and hence of the $\lambda$
using $\mu$SR technique \cite{Sonier}. 
Unfortunately, single crystals of SLCO are not available at present.
Concerning the pairing symmetry in SLCO based on other experiments, we 
note that the recent tunneling experiments suggest strong-coupling $s$-wave
pairing in SLCO \cite{Chen}. On the other hand, NMR spin-lattice relaxation
and Knight shift measurements were found to be more consistent with
the line-nodes gap \cite{Williams}. It remains to be understood why different 
experimental techniques provide controversial results concerning the pairing
symmetry in SLCO.

In summary, we performed TF-$\mu$SR measurements of the 
in-plane penetration depth $\lambda_{ab}$ in  $n$-type ILS 
Sr$_{0.9}$La$_{0.1}$CuO$_{2}$. Absence of the magnetic rare-earth 
elements in this compound allowed to measure for the first time the 
absolute value of  $\lambda_{ab}(0)$ in  $n$-type HTS using $\mu$SR. We 
found $\lambda_{ab}(0)$=116(2) nm. The zero-temperature depolarization 
rate $\sigma(0)\propto 1/\lambda^{2}(0)$= 4.6(1) $\mu$s$^{-1}$ is
more than four times higher than expected from the Uemura line. 
Therefore this $n$-type HTS does not follow the Uemura relation in
contrast to $p$-type HTS. We also performed ZF-$\mu$SR experiments and 
found no evidence of magnetic order in SLCO. This indicates the 
competitive character of AF order and superconductivity in $n$-type 
cuprates in agreement with the recent neutron scattering experiments 
\cite{Fujita}.

This work was supported by the Swiss National Science Foundation and 
by the NCCR program {\it Materials with Novel Electronic Properties} 
(MANEP) sponsored by the Swiss National Science Foundation.

\end{document}